\documentclass[sigconf]{acmart}

\usepackage{url}
\usepackage{subfig}
\usepackage{balance}

\def\BibTeX{{\rm B\kern-.05em{\sc i\kern-.025em b}\kern-.08emT\kern-.1667em\lower.7ex\hbox{E}\kern-.125emX}}
    
\copyrightyear{2020}
\acmYear{2020}
\acmConference[MobiCom 2020]{26th Annual International Conference on
Mobile Computing and Networking}{September 21--25, 2020}{London, United Kingdom}
\acmBooktitle{The 26th Annual International Conference on
Mobile Computing and Networking (MobiCom 2020),
September 21--25, 2020, London, United Kingdom}
\acmPrice{15.00}
\acmDOI{10.1145/xxxxxxx.xxxxxxx}
\acmISBN{978-x-xxxx-xxxx-x/xx/xx}

\begin{document}

%
% The "title" command has an optional parameter, allowing the author to define a "short title" to be used in page headers.
\title[Private 4G/5G Network]{Demo: Slicing-Enabled Private 4G/5G Network for \\Industrial Wireless Applications}

%
% The "author" command and its associated commands are used to define the authors and their affiliations.
% Of note is the shared affiliation of the first two authors, and the "authornote" and "authornotemark" commands
% used to denote shared contribution to the research.

%\author{Adnan Aijaz}
%\affiliation{%
% \institution{Toshiba Research Europe Ltd.}
% \streetaddress{32 Queen Square}
% \city{Bristol}
% \country{U.K.}}
% \email{adnan.aijaz@toshiba-trel.com}
% 

\author{Jaya Thota, Adnan Aijaz}
\affiliation{%
 \institution{Bristol Research and Innovation Laboratory, Toshiba Europe Ltd., Bristol, United Kingdom}
 \streetaddress{32 Queen Square}
 %\city{Bristol}
 %\country{U.K.}}
 }
 \email{{jaya.thota, adnan.aijaz}@toshiba-bril.com}

%
% By default, the full list of authors will be used in the page headers. Often, this list is too long, and will overlap
% other information printed in the page headers. This command allows the author to define a more concise list
% of authors' names for this purpose.

\renewcommand{\shortauthors}{Thota and Aijaz}

%
% The abstract is a short summary of the work to be presented in the article.
\begin{abstract}
Private deployments of 4G and 5G networks in industrial environments are beneficial from various aspects. Private 4G/5G networks typically face the challenge of supporting heterogeneous industrial applications. This technology demonstration highlights the importance of network slicing in private 4G/5G networks. It shows that network slicing is crucial for performance guarantees in multi-service co-existence scenarios. With network slicing, our private 4G/5G network successfully supports closed-loop control, event-driven control and video streaming applications. 
\end{abstract}

\maketitle

\section{Introduction}
The  5G mobile/cellular technology is widely recognized as the frontrunner among the candidate technologies for the envisioned transformation of industrial systems \cite{TI_PIEEE, fut_ind_comm}. Operation of \emph{private} (non-public) networks in industrial environments is promising to fully unleash the potential of 5G. Private deployments offer dedicated coverage, exclusive use of capacity and the capability of a customized service while providing complete control over the network \cite{pvt_5G}. Private 4G networks have already been deployed across various industrial domains including mining, warehousing and utilities \cite{harbor_pvt}.  Recent regulatory initiatives on opening of \emph{shared licensed spectrum} empower industrial stakeholders to deploy their own local networks with dedicated equipment and settings. 

%\footnote{Notable examples include the 3.7 -- 3.8 GHz band in Germany, the 3.8 -- 4.2 GHz band in the U.K. and the 3.5 GHz band in the U.S. }

Due to the increasingly heterogeneous nature of industrial communication, network slicing \emph{within} a private network becomes particularly important. Network slicing provides the capabilities of traffic isolation and application-specific resource management which are crucial for strict performance guarantees in multi-service co-existence scenarios. Slicing of the radio access network (RAN) is a key element of network slicing \cite{3gpp.38.300}. Realizing RAN slicing through a gateway deployed inside a private network provides various benefits \cite{pvt_5G}. Gateway-level slicing approach is closely aligned with the software-defined networking (SDN) paradigm for RAN. 

The main focus of this demonstration is a slicing-enabled private 4G/5G network for industrial communication. It demonstrates the viability of a slicing-enabled private 4G/5G network as a single wireless interface for versatile industrial applications. It shows the feasibility of gateway-level RAN slicing in private networks. It also shows the capability of operation in the recently-opened 3.8 -- 4.2 GHz shared licensed spectrum for private 5G in the U.K. \cite{ofcom_lic}. 
%Last, but not least, it exhibits the potential of open innovation for private networks.  

\section{Demonstration Overview}
\begin{figure}
\centering
\includegraphics[scale=0.28]{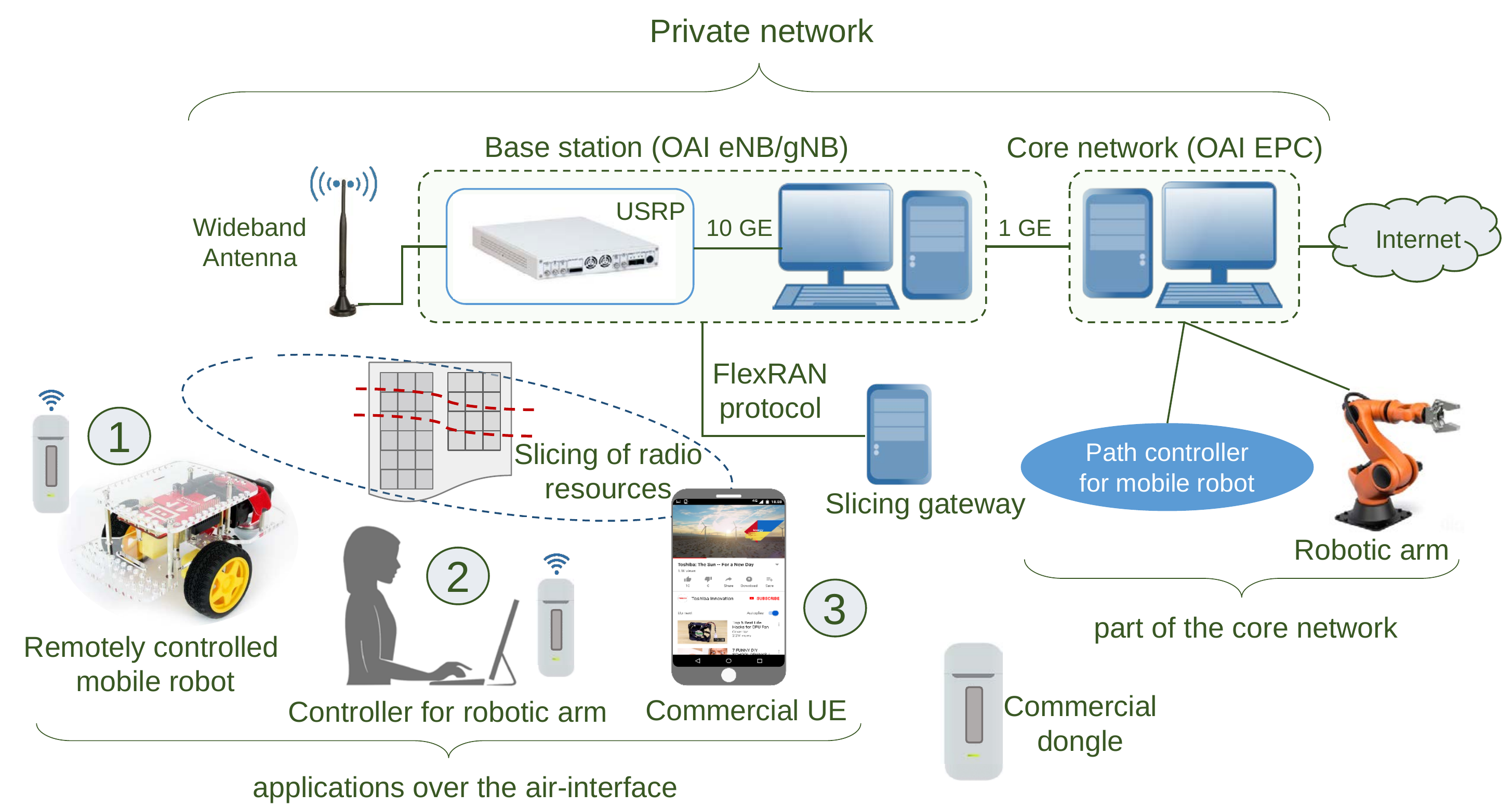}
\caption{Illustration of the demonstration scenario.}
\label{scenario}
\vspace{-2em}
\end{figure}

The demonstration scenario is illustrated in \figurename~\ref{scenario}. Our private network has three core components: a RAN, a core network (CN) and a slicing gateway. We consider three different applications on the private network. The first application is \emph{closed-loop control} wherein a path controller located in the CN \emph{remotely} drives a mobile robot on a pre-defined path. There is bi-directional exchange of command and feedback messages between the path controller and the mobile robot. Such bi-directional traffic must be handled with low and deterministic latency. The second application is \emph{event-driven control} wherein a pre-programmed robotic arm located in the CN receives actuating commands from a human operator on the air-interface. This creates uplink only traffic which must be transmitted with low latency. The third application is \emph{video streaming} wherein a commercial UE is streaming high-definition (HD) video from the Internet. This creates bandwidth-hungry traffic in the downlink. Our demonstration shows successful co-existence of these very different applications with slicing of radio resources. 

\section{Design and Implementation}
\subsection{Private Network}
In 4G mode, the base station front-end is deployed using commercial off-the-shelf (COTS) Ettus X310 software-defined radio (SDR) platform connected via a 10 Gigabit Ethernet (GE) interface to a server that is characterized by  Intel Core i7-8700 12-core (3.2 GHz) processor and 16 GB RAM. A wideband (850 MHz -- 6.5 GHz) directional antenna with a gain of 5-6 dBi is used at the front-end. The base station server (eNB) is running  "Ubuntu 16.04" operating system with "Linux kernel" release 4.15.0-041500-low latency. It is connected to the CN and the slicing gateway via 1 GE interfaces. The base station operates in 2.6 GHz band with 10 MHz bandwidth. We use low transmit power in demonstration scenario as per the regulatory requirements for license-free operation in this licensed band.
The CN and the slicing gateway run on different servers with similar specifications as the base station server with the exception of   low-latency kernel. The private network implements standard-compliant open source software stack for base station and CN servers based on OpenAirInterface (OAI) \citep{OAI_5G, OAI_LTE}. Commercial 4G dongles and handsets are used as user equipments (UEs) to communicate over the private network. All UEs have Open Cells SIM cards programmed using  parameters in OAI home subscriber service (HSS) database with separate international mobile subscriber identity (IMSI). The slicing gateway is based on Mosaic5G service platforms \cite{mosaic_5G}.
In the non-standalone (NSA) mode of 5G New Radio (NR), the RAN includes a second base station server (gNB) connected to a second Ettus X310 front-end (capable of supporting higher bandwidths). The gNB and the eNB are connected via a 1 GE interface.

%\subsection{Applications}
For closed-loop control, we use a GoPiGo3 robotic car stacked on a Raspberry Pi with ATMEGA328 microcontroller. The GoPiGo3  robot is a differential drive system with two driving wheels and one caster wheel. The microcontroller sends, receives and executes commands sent by the Raspberry Pi to perform motor control. A wireless dongle connected to the Raspberry Pi USB interface communicates with the path controller \cite{demo_infocom} in the CN. The path controller remotely drives the robot along a pre-defined trajectory by controlling the speed of driving wheels. This reflects a centrally-controlled mobile platform in warehousing and logistics. The path controller periodically receives motor encoder values for driving wheels as feedback (in uplink) via the wireless dongle. It transmits updated parameters (in  downlink), based on robot's kinematics and path deviation error. 
For event-driven control, we use a DOBOT magician robotic arm that is connected to the CN server over a USB interface. The robotic arm tool head is equipped with a gripper or a suction cup to pick and place wooden blocks, imitating loading/unloading of objects on a production line. The event-driven control commands to the robotic arm are issued by a human via a laptop connected to a wireless dongle. A Kinetic robot operating system (ROS) API is written to control the robotic arm such that the CN is the master node and the laptop is the slave node. The actuating commands which are sent over the private network via the wireless dongle are translated by the CN to ROS commands for actions performed by the robotic arm. 
For video streaming, a commercial handset running YouTube application is used.

\begin{figure}
\centering
\includegraphics[scale=0.48]{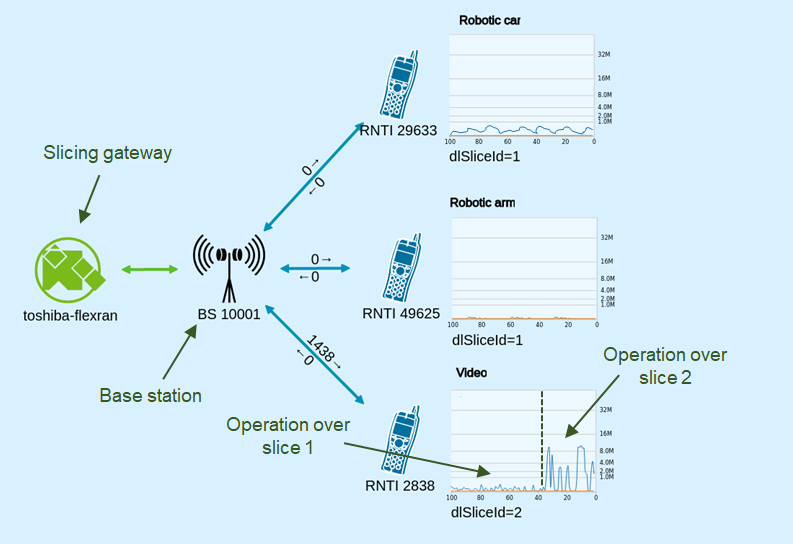}
\caption{Real-time RAN connectivity map. }
\label{res_slicing}
\vspace{-2em}
\end{figure}

\subsection{Radio Resource Slicing and Results}
Our radio resource slicing strategy is based on FlexRAN \cite{flexRAN} which is a flexible and programmable software-defined RAN platform.  The FlexRAN real-time master controller running on the slicing gateway server is connected to the FlexRAN agent on the base station server via southbound API (OpenFlow). The FlexRAN controller also provides a REST northbound API for RAN control and management. Using these APIs, we have created two radio slices: slice 1 for control applications and slice 2 for high data rate applications. Each slice is defined in downlink and uplink with a slice ID, slice label, percentage of allocated resources, slice priority, and availability of resource blocks (high and low). Slice 1 is defined with only \(5\%\) of radio resources but with higher availability of resource blocks, whereas slice 2 is defined with the remaining \(95\%\) of radio resources but with lower availability of resource blocks. Our slicing strategy is capable of dynamically adjusting allocated resources for each slice depending on resource utilization and utility. It also provides dynamic relocation of UEs in order to fulfil service requirements.

%\subsection{Results}
We consider two scenarios in the demonstration. In the \emph{baseline} scenario, no slices are instantiated and the available radio resources are shared among the UEs such that control applications are prioritized. In the \emph{slicing-enabled} scenario, radio resources are allocated as per the aforementioned slicing strategy. \figurename~\ref{res_slicing} shows real-time RAN behavior based on drone application in Mosaic5G Store platform. Initially, all the UEs are connected to slice 1 with the graphs showing throughput (in Mbps). This reflects the baseline scenario. Owing to prioritized allocation and small data packets,  control applications run smoothly over slice 1. However, the allocated resources are insufficient for video streaming. Hence, the video keeps buffering and freezing. The slicing controller relocates the video UE (with RNTI 2838), using the REST API, from slice 1 to slice 2 which is optimized for high data applications. After relocation, the video stream runs smoothly as shown by the throughput graph. This reflects the slicing-enabled scenario which provides slice isolation capabilities for successful co-existence of different applications over the air-interface. 

We have also successfully tested operation in 5G-NR NSA mode (30 kHz sub-carrier spacing, 80 MHz bandwidth), currently supported by OAI, after obtaining a license for the recently-opened 3.8 - 4.2 GHz band for private networks. A complete demonstration in 5G NSA mode is not possible as commercial devices for this band are not available yet.

\section{Remarks} \label{sect_cr}
This demonstration reveals the importance of network slicing within  private 4G/5G networks. It shows the effectiveness of real-time on-the-fly control delegation and management of radio resources for providing performance guarantees in multi-service co-existence scenarios. It also shows one of the first slicing-enabled 5G-ready private network, operating in shared licensed spectrum, for versatile industrial wireless applications. Finally, it highlights the potential of COTS hardware and open source software for private 4G/5G networks. A short video of the demonstration is available at https://tinyurl.com/y862wfp8. Our 5G-NR NSA setup is shown in https://tinyurl.com/yaj4o2xl.

%
%The ``\verb|sigchi-a|'' template style (available only in \LaTeX\ and not in Word) produces a landscape-orientation formatted article, with a wide left margin. Three environments are available for use with the ``\verb|sigchi-a|'' template style, and produce formatted output in the margin:
%\begin{itemize}
%\item {\verb|sidebar|}:  Place formatted text in the margin.
%\item {\verb|marginfigure|}: Place a figure in the margin.
%\item {\verb|margintable|}: Place a table in the margin.
%\end{itemize}

%
% The acknowledgments section is defined using the "acks" environment (and NOT an unnumbered section). This ensures
% the proper identification of the section in the article metadata, and the consistent spelling of the heading.
%\begin{acks}
%To Robert, for the bagels and explaining CMYK and color spaces.
%\end{acks}

%
% The next two lines define the bibliography style to be used, and the bibliography file.
\balance
\bibliographystyle{ACM-Reference-Format}
%\bibliography{sample-base}
\bibliography{demo_ref}
%\printbibliography{\textsf{\textsf{GALLOP}}_bib.bib}
% If your work has an appendix, this is the place to put it.
\appendix

\end{document}